\documentclass[conference]{IEEEtran}
\IEEEoverridecommandlockouts
\usepackage{cite}
\usepackage{amsmath,amssymb,amsfonts}
\usepackage{algorithmic}
\usepackage{graphicx}
\usepackage{subcaption}
\usepackage{textcomp}
\usepackage{xcolor}
\usepackage[bookmarks=false]{hyperref}
\hypersetup{nolinks=true}
\usepackage{booktabs}
\usepackage{tabularx}
\usepackage{listings}
\usepackage{color}
\usepackage[autostyle=true]{csquotes}
\newcommand{\sysml}[1]{\texttt{#1}}
\newcommand{\fpb}[1]{#1}
\newcommand{\codein}[1]{\texttt{#1}}

\lstdefinestyle{XMLStyle}{
    language=XML,
    basicstyle=\tiny\ttfamily,
    numbers=left,
    numberstyle=\tiny,
    numbersep=5pt,
    keywordstyle=\color{blue},
    morecomment=[l]{\#},
    commentstyle=\color{black}\bfseries,
    stringstyle=\color{purple},
    frame=tb,
    breaklines=true,
    showstringspaces=false,
    morekeywords={element, attribute, value, state, references, identification, characteristics, assignments, assigned, flows, flow, exit, entry, process, systemLimit, states, references}
}

\lstdefinestyle{OCLStyle}{
    basicstyle={\tiny\ttfamily},
    numbers={left},
    numberstyle={\tiny},
    numbersep={2pt},
    keywordstyle = {\color{blue}},
    keywordstyle = [2]{\color{black}\bfseries},
    morecomment=[l]{\#},
    commentstyle=\color{black}\bfseries,
    stringstyle=\color{purple},
    morestring=*[d]{'},
    frame={tb},
    breaklines={true},
    showstringspaces={false},
    otherkeywords = {context, inv},
    morekeywords={and, not, or},
    morekeywords = [2]{context, inv}
}

\usepackage[nolist]{acronym} 
\usepackage{comment} 
\def\BibTeX{{\rm B\kern-.05em{\sc i\kern-.025em b}\kern-.08em
    T\kern-.1667em\lower.7ex\hbox{E}\kern-.125emX}}
\begin{document}
\begin{acronym}[ECU]
\acro{aml}[AML]{Automation Markup Language}
\acro{dsml}[DSML]{Domain-Specific Modeling Language}
\acroplural{dsml}[DSMLs]{Domain-Specific Modeling Languages}
\acro{fas}[FAS]{Functional Architecture for Systems}
\acro{fpd}[FPD]{Formalised Process Description}
\acroplural{fpd}[FPDs]{Formalised Process Descriptions}
\acro{ibd}[IBD]{Internal Block Diagram}
\acroplural{ibd}[IBDs]{Internal Block Diagrams}
\acro{json}[JSON]{Java Script Object Notation}
\acro{mbse}[MBSE]{model-based systems engineering}
\acro{msosa}[MSoSA]{Magic Systems of Systems Architect}
\acro{ppr}[PPR]{product, process, and resource}
\acro{sysml}[SysML]{Systems Modeling Language}
\acro{tags}[Tags]{Tagged Values}
\acro{ocl}[OCL]{Object Constraint Language}
\acro{odp}[ODP]{Ontology Design Pattern}
\acro{omg}[OMG]{Object Management Group}
\acro{owl}[OWL]{Web Ontology Language}
\acro{uml}[UML]{Unified Modeling Language}
\acro{vtl}[VTL]{Velocity Template Language}
\acro{xmi}[XMI]{XML Metadata Interchange}
\acro{xml}[XML]{Extensible Markup Language}

\acro{soi}[SoI]{System of Interest}
\acro{bdd}[\textbf{[bdd]}]{SysML Block Definition Diagram}
\acro{ibd}[\textbf{[ibd]}]{SysML Internal Block Diagram}
\acro{uc}[\textbf{[uc]}]{SysML Use Case Diagram}
\acro{act}[\textbf{[act]}]{SysML Activity Diagram}
\acro{pkg}[\textbf{[pkg]}]{SysML Package Diagram}
\acro{stm}[\textbf{[stm]}]{SysML State Machine Diagram}

\acro{cps}[CPS]{cyber-physical system}
\acroplural{cps}[CPSs]{cyber-physical systems}
\acro{kpi}[KPI]{key performance indicator}
\acroplural{kpi}[KPIs]{key performance indicators}
\acro{csmop}[CSMOP]{constraint satisfaction and multi-criteria objective problem}

\acro{fwb}[FWB]{Functional Whitebox}
\acro{rflt}[RFLT]{Requirement-Functional-Logical-Technical}
\acro{spes}[SPES]{Software Platform Embedded Systems}
\acro{pddl}[PDDL]{Planning Domain Definition Language}
\acro{hse}[HSE]{Health-Safety-Environment}


\acro{ad}[AD]{Activity Diagram}
\acroplural{ad}[ADs]{Activity Diagrams}

\end{acronym}

\title{A SysML Profile for the Standardized Description of Processes during System Development
\thanks{This research in the iMOD project is funded by dtec.bw – Digitalization and Technology Research Center of the Bundeswehr. dtec.bw is funded by the European Union – NextGenerationEU.}
}
\author{
\IEEEauthorblockN{
Lasse Beers\IEEEauthorrefmark{1},
Hamied Nabizada\IEEEauthorrefmark{1},
Maximilian Weigand\IEEEauthorrefmark{1},
Felix Gehlhoff\IEEEauthorrefmark{1},
Alexander Fay\IEEEauthorrefmark{1}\IEEEauthorrefmark{2}}
\IEEEauthorblockA{\\ \IEEEauthorrefmark{1}Institute of Automation Technology\\
Helmut Schmidt University, Hamburg, Germany\\
\{lasse.beers, hamied.nabizada, maximilian.weigand, felix.gehlhoff, alexander.fay\}@hsu-hh.de\\
\\
\IEEEauthorrefmark{2}Chair of Automation Technology\\
Ruhr University, Bochum, Germany\\
alexander.fay@rub.de}}

\maketitle

\begin{abstract}
A key aspect in creating models of production systems with the use of \ac{mbse} lies in the description of system functions. 
These functions should be described in a clear and standardized manner. 

The VDI/VDE 3682 standard for \ac{fpd} provides a simple and easily understandable representation of processes. 
These processes can be conceptualized as functions within the system model, making the \ac{fpd} particularly well-suited for the standardized representation of the required functions.
Hence, this contribution focuses on the development of a \ac{dsml} that facilitates the integration of VDI/VDE 3682 into the \ac{sysml}.
The presented approach not only extends classical \ac{sysml} with domain-specific requirements but also facilitates model verification through constraints modeled in \ac{ocl}. 
Additionally, it enables automatic serialization of process descriptions into the \ac{xml} using the \ac{vtl}. 
This serialization enables the use of process modeling in applications outside of \ac{mbse}. 
The approach was validated using an collar screwing use case in the major component assembly in aircraft production. 
\end{abstract}

\begin{IEEEkeywords}
SysML Profile, Model-Based Systems Engineering, Domain-Specific Modeling Language, VDI/VDE 3682, Formalised Process Description
\end{IEEEkeywords}

\acresetall
\section{Introduction} 
Approaches of \ac{mbse} enable system architects to respond more quickly and effectively to the numerous changes in the requirements that occur during the development process \cite{d2017systems}. 
Consequently, modeling and management of production resources and planning of resource allocation are becoming increasingly important for manufacturing companies \cite{sanfilippo2018modeling}.

In the realm of \ac{mbse}, it is a well-established practice to investigate the required functions for the system model and capture them in a Blackbox and subsequently providing detailed specifications within a Whitebox \cite{pohl2012model}. 
The Whitebox includes the definition of the inputs and outputs of functions. 
The way these inputs and outputs are exchanged in between functions as well as the sequential order of functions are called functional relationships and are also part of the Whitebox definition. 
The key experts for defining functional relationships are typically process engineers. 
Although these engineers possess the necessary expertise, they are often familiar only with their own descriptive tools of process modeling and not with the widely used modeling language \ac{sysml} and its associated software tools in the context of \ac{mbse}. 

One such descriptive tool for process modeling is the \ac{fpd} according to VDI/VDE~3682~\cite{VDIVDEGesellschaftMessundAutomatisierungstechnik.05.2015}, which is also easily understandable for other engineers and even non-technical personnel. 
The \ac{fpd} precisely defines relationships between in- and output of interrelated processes. 
These processes can be understood as functions in the system model and therefore the \ac{fpd} is highly suitable for the modeling of the Whitebox. 
At the same time, by considering the incoming and outgoing products, as well as the resources required for process execution, it establishes a solid foundation for addressing the prevalent connection between \ac{ppr} in the manufacturing industry during the modeling of production systems~\cite{gehlhoff2022challenges}.   

The aim of this paper is, therefore, to demonstrate how modeling according to the VDI/VDE 3682 can be achieved within the tools in the field of \ac{mbse}. 
For this purpose, a \ac{sysml} profile has been developed and is presented in this contribution as a \ac{dsml} to model system functions. 
This profile can be utilized in modeling system functions to depict the desired behavior and corresponding functional inputs and outputs. 
Simultaneously, it reduces the complexity of modeling the functional Whitebox while increasing the standardization of the system model.  
Modeling according to standards enhances understanding and acceptance among participants while facilitating the establishment of universally applicable formal rules, simplifying model verification.

The following sections are structured as follows: 
In Section~\ref{sec:fundamentals}, the fundamentals of this approach are explained. 
Section~\ref{sec:sota} provides an overview of existing approaches to applications using the \ac{fpd} and modeling system functions in \ac{mbse}. 
In Section~\ref{sec:method}, the \ac{dsml} for the VDI/VDE 3682 is introduced, and in Section~\ref{sec:applicationexample}, it is applied using an example from aircraft production. 
The paper concludes with a summary and outlook in Section~\ref{sec:summaryoutlook}. 

\section{Fundamentals}\label{sec:Related Work}
\label{sec:fundamentals}
\subsection{VDI/VDE 3682 Formalised Process Description}
The VDI/VDE 3682 standard \cite{VDIVDEGesellschaftMessundAutomatisierungstechnik.05.2015} describes the conception and use of the \ac{fpd}. 
It is a universal tool for the description of all types of processes, both technical and non-technical. 
All process-related information can be clearly and systematically communicated, reused, and progressively detailed throughout the entire lifecycle of an automated system. 

\begin{figure}[htp]
    \centering
    \includegraphics[width=0.425\textwidth]{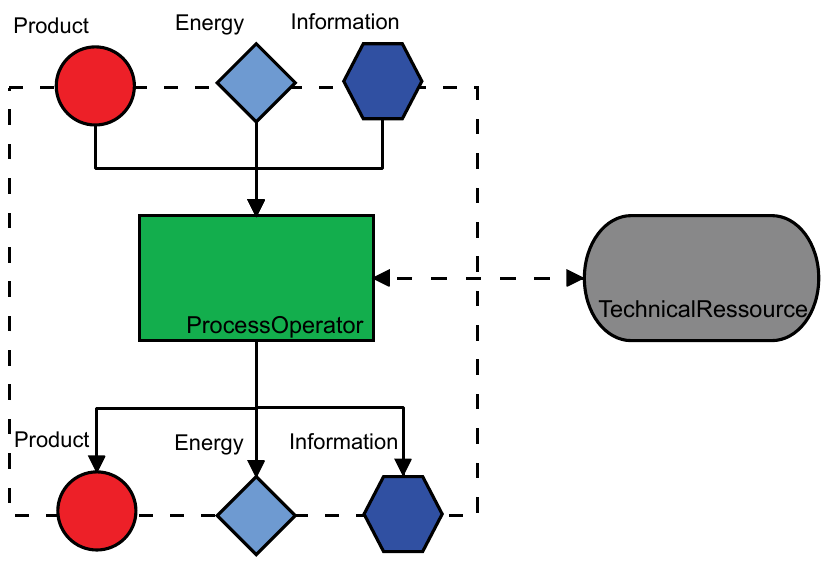}
    \caption{Graphical Elements of the \ac{fpd}}
    \label{fig:fdp_graphix}
\end{figure}

For the graphical modeling of a process, the standard defines a set of permissible symbols and their relationships. 
\fpb{State}-describing symbols (\fpb{Product}, \fpb{Energy}, \fpb{Information}) are connected to a \fpb{Process Operator} by directed edges (\fpb{flow} connections). 
The \fpb{Process Operator} describes a technology-neutral transformation of a state ante to a state post. 
The technical realization of this \fpb{Process Operator} is represented by a \fpb{usage} relationship between the \fpb{Process Operator} and a \fpb{Technical Resource}.  
To delineate the process from its external environment, a \fpb{System Boundary} is defined around the modeled process. 
The elements \fpb{Product}, \fpb{Energy}, or \fpb{Information} located on this \fpb{System Boundary} can be understood as inputs or outputs and thus indicate the interactions of the system with its environment. 
Within the \fpb{System Boundary}, individual process steps can be further specified through the decomposition of the respective \fpb{Process Operators}.  
A small abstract example featuring the graphical elements is outlined in \autoref{fig:fdp_graphix}.

In addition, both parallel and alternative process flows can be represented without expanding the symbol set. 
In graphical modeling, a parallel flow is represented by a partially shared flow, while alternative flows are represented by straight flows. 

\begin{figure}[htp]
    \centering
    \includegraphics[width=0.45\textwidth]{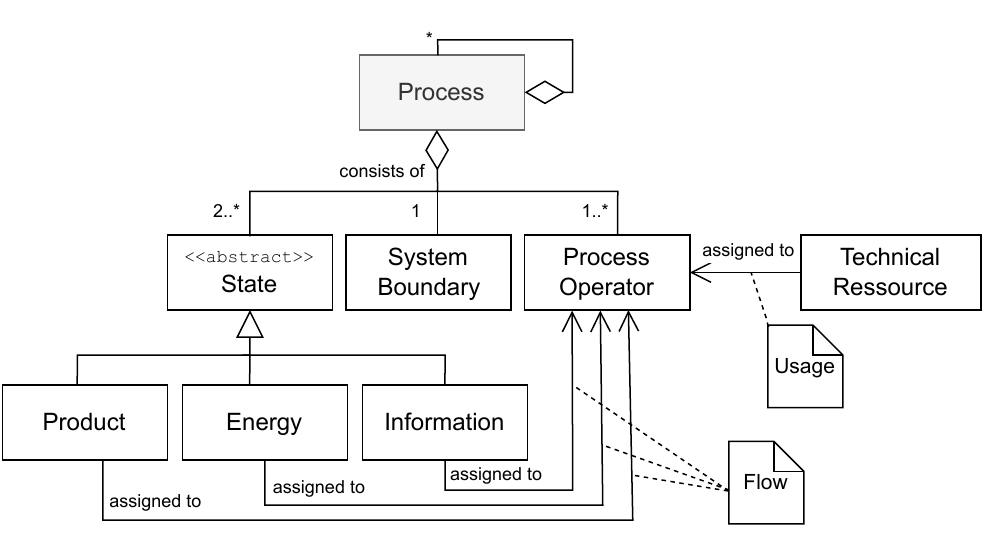}
    \caption{Class Diagram of the \ac{fpd}. Adapted from \cite{  VDIVDEGesellschaftMessundAutomatisierungstechnik.05.2015.part2}.}
    \label{fig:fdp_classdia}
\end{figure}

The graphical notation of the \ac{fpd} is supported by an underlying object-oriented information model \cite{  VDIVDEGesellschaftMessundAutomatisierungstechnik.05.2015.part2}.
The class diagram of this information model is shown in \ac{uml} notation in \autoref{fig:fdp_classdia} and forms the basis for the developed profile of this contribution, which is introduced in Section~\ref{sec:method}.

Furthermore, the standard VDI/VDE 3682 outlines how the identification of individual elements and the description of their features can be modeled. 
Identification includes, for example, a unique ID, a short name, and a long name \cite{VDIVDEGesellschaftMessundAutomatisierungstechnik.05.2015.part2}.

\subsection{Domain-Specific Modeling Language (DSML)}
\acp{dsml} enable developers to reconstruct and apply domain-specific concepts precisely in a suitable modeling language. 
The use of such languages contributes to enhancing the quality of modeling, as inconsistent modeling can be avoided to some extent through the clear syntax and semantics of \acp{dsml}\cite{frank2013domain}. 

\ac{uml} offers a wide range of possibilities to model knowledge in the form of various elements and diagrams. 
However, the variety offered is often not sufficient for modeling domain-specific knowledge. 
In order to meet the domain-specific modeling challenges, a lightweight or a heavyweight \ac{dsml} can be developed. 
Modifying the existing \ac{uml} metamodel or even developing an entirely new metamodel is referred to as a heavyweight approach \cite{lagarde2008leveraging}. 
If the existing \ac{uml} metamodel is taken as the base without any modifications and is extended with domain-specific requirements or notations, it is referred to as a lightweight approach~\cite{seidl2012uml}. 

Specially for lightweight approaches, the \ac{uml} profiling mechanism was developed, allowing an extension of the metamodel. 
The most well-known example of such a \ac{uml} profile is \ac{sysml}. 
It extends \ac{uml} with additional elements and diagrams and has become one of the most established languages for modeling complex systems  \cite{seidl2012uml}. 
A key advantage of using this mechanism is the standardization and the associated reusability of a profile.  
Through the standardized exchange format called \ac{xmi} established by the Object Management Group, a developed profile can be easily imported or exported, making it accessible to other users. 
The central component of the profile is the stereotype.  
 It is a distinct metaclass within \ac{uml} that can extend or restrict other existing metaclasses by adding additional meta-attributes. 
A stereotype can also inherit properties from multiple metaclasses or other stereotypes. 
Furthermore, the visual appearance of the altered stereotype can be modified. 
The extension of a stereotype through meta-attributes is implemented by creating \ac{tags}. 
\ac{tags} are attributes specifically created for a stereotype. 

The simplest form of \ac{tags} are regular attributes.  
When added, every element of the modified stereotype acquires this property. 
However, more complex \ac{tags} can also be developed.  
A commonly used example is the Derived Property, which is a characteristic whose value is defined by its relationship with other elements. 
The value is dynamically calculated, ensuring it automatically updates during model adjustments~\cite{nomagic.2015}. 
Derived Properties can be defined by a simple relationship, a linked meta chain, or by other types of code. 
These are implemented within the element \sysml{Customization}. 
However, Customization also provides a way to limit the use of a stereotype.
This includes the definition of Connection Rules or Model Initialization. 
Connection Rules, such as \sysml{TypeOfSource}/\sysml{TypeOfTarget}, restrict the scope of elements that can be the source of a relationship to elements belonging to a certain stereotype. An example of this is that a usage relationship (the dotted line between the process operator and the technical resource in \autoref{fig:fdp_graphix}) can only be used to connect the corresponding stereotypes. 
Similarly, visibility and usage within other elements can be manipulated. 
However, these limitations defined through the customization are not sufficient to impose all possible restrictions. 

Ths fact leads us to another essential tool for restricting a profile which is called Constraints. 
These can be developed independently of the profile and later added to one or more stereotypes.  
Constraints can be modeled in \ac{ocl} or in programming languages like Java. 
Within the Constraint, complex relationships can be defined~\cite{bousse2012aligning}. 
For example, it can be defined that each FBD diagram must have at least one input state and one output state.
These states are recognized by being on the system boundary. 

The development of new diagrams, specifically tailored to the domain of \ac{dsml}, is also possible. 
Completely new diagrams can be developed, or existing \ac{uml}/\ac{sysml} diagrams can be used as a foundation, which is then extended or restricted according to the specific needs.  
For this purpose, the newly developed stereotypes are defined as usable elements for the corresponding diagram. 

Furthermore, various software tools provide additional functionalities to support the individual creation of a \ac{dsml} \cite{cabot2012object}. 
For example, the software tool \enquote{\ac{msosa}}, developed by Dassault Systèmes, provides advanced customization possibilities, such as implementing custom logic with Java code.  

\section{State of the art}
\label{sec:sota}
In this section, approaches from VDI/VDE 3682 that utilize the \ac{fpd} to create various model types are presented. 
Concurrently, the current status of potential serialization efforts is outlined. 
The second part of this section introduces various \ac{mbse} approaches that encompass the modeling of system functions and the Blackbox and Whitebox.  

\subsection{Usage and Serialization of the VDI/VDE 3682}
\label{sec:sota_3682approaches}
Since the publication of the VDI/VDE 3682, it has been gaining increasing significance due to its diverse applications. 
In \cite{hildebrandt2018ontology}, an Ontology Design Pattern is introduced to map the information model of the standard, offering a reusable pattern for process descriptions using ontologies in \ac{owl}. 
A possible mapping of the \ac{fpd} into \ac{aml}  \cite{drath2021automationml} was presented in \cite{jager2012durchgangige}. 
The \ac{fpd} is also chosen as a suitable means of description in capability models (e.g., in \cite{kocher2020formal} or \cite{csspaper}) and digital process twins \cite{caesar2020information}. 
The authors of \cite{tj2023} use the \ac{fpd} as a starting point for the mathematical description of interdependencies among process parameters.
In \cite{kathrein2019meta}, a formal metamodel was developed based on the \ac{fpd}, enabling consistency checks. 
The authors of \cite{novak2022digitalized} use the \ac{fpd} as a basis for an intelligent production process planning algorithm. 
An open-source web-based tool\footnote{ publicly available at https://demo.fpbjs.net} facilitates standard-compliant process modeling~\cite{nabizada2020offenes}. 
This tool employs a custom serialization format using JSON for storing process descriptions. 

The serialization in XML format announced for part 3 of the VDI/VDE 3682 standard has not been standardized yet. 
However, a proposal for such serialization is provided in \cite{nabizada2022vorschlag}. 
The serialization of the process descriptions described in Section~\ref{sec:automatedcodegeneration} follows the proposal in \cite{nabizada2022vorschlag}. 
To the best of our knowledge, there is currently no approach that represents the \ac{fpd} as a \ac{dsml} based on \ac{sysml}. 
\subsection{Approaches for Modeling System Functions with SysML}

\ac{sysml} stands out as a widely adopted graphical modeling language, establishing itself as the de-facto standard within the realm of \ac{mbse} . 
Renowned for its versatility, \ac{sysml} provides an extensive range of capabilities, facilitating the specification, analysis, design and verification of intricate systems \cite{ObjectManagementGroup.2019}. 

The \ac{spes} Modeling Framework was created within the context of various research endeavors, including SPES2020~\cite{pohl2012model}, SPES XT \cite{pohl2016advanced}, and CreSt \cite{bohm2021model}. 
Its objective is to provide language-independent modeling methods for the design of embedded systems and software systems. 
These methods have also been implemented in various approaches, such as those presented in \cite{hayward2022sysml} and \cite{beers2023mbse}, utilizing the \ac{sysml} modeling language.
The framework adheres to an iterative process encompassing the delineation of distinct layers and viewpoints. 
To capture the functional viewpoint, it proposes a description utilizing Blackbox and Whitebox functions.
This approach aims to mitigate complexity by hierarchically structuring functionality from the user's perspective, overseeing functional interactions, and comprehending functional relationships through the capturing and analyzing of interactions between different (sub-)functionalities.
Notably, the functions within the functional Whitebox model must collectively exhibit a behavior identical to that specified by the user function in the functional Blackbox model \cite{pohl2012model}.
 
Based on the knowledge and methods developed within in the context of SPES and CreSt, Hayward et al. introduce a function-centered approach for describing collaborative cyber-physical systems, employing the \ac{sysml} \cite{hayward2022sysml}.
The authors emphasize the importance of modeling Blackbox and Whitebox for the function-centric view. 
Blackbox functions are modeled as a stereotype of the metaclass Activity, representing the functional structure of the system group. 
Whitebox stereotypes are then utilized to specify the Blackbox, defining the sequential process and assignment to roles. 
However, the approach does not involve modeling inputs and outputs for the definition of Blackbox and Whitebox. 
The correctness of the modeling is not guided by corresponding rules or constraints.

Welkiens et al. propose to design the functions of a system using the \ac{fas} method \cite{weilkiens2022model}. 
This method involves grouping the activities of the use case through the identification of functional groups. 
Subsequently, taking into account both functional and non-functional requirements, the functional architecture is modeled using a SysML Internal Block Diagram. 
The functional groups are related and connected through corresponding ports. 
However, this approach quickly leads to a cluttered and complex representation of system functions, making it understandable and implementable only by modeling experts. 
Additionally, the lack of standardization introduces a considerable degree of interpretation in the modeling process. 

Binder et al. outline in their approach \cite{binder2023towards} the necessity of utilizing the \ac{ppr} concept in the realm of Systems Engineering, particularly when modeling a production system.  
The authors have developed a \ac{uml} metamodel for representing the \ac{ppr} context and employ process descriptions at the functional level. 
However, the approach only considers products as inputs for processes and lacks reliance on any standard. 
Furthermore, no verification mechanisms are introduced. 

The modeling approach presented in \cite{beers2023mbse}, which is also aligned with the SPES method, incorporates the modeling of both Blackbox and Whitebox aspects within the definition of the functional viewpoint.
In this approach, the \ac{fpd} is used as a basis for modeling. 
The inputs and outputs of the functions are modeled through \fpb{State}-describing elements prescribed in the standard, namely \fpb{Information}, \fpb{Product}, and \fpb{Energy}. 
However, simple \ac{sysml} elements are used for modeling, which prevents guided modeling due to the limitations in the use or verification of the models.  
\newline

In summary, none of the presented approaches in \ac{mbse} allows for the comprehensive modeling of the \ac{ppr} concept according to a standard. 
This gap is addressed by the design of a \ac{sysml} profile for the VDI/VDE 3682 standard\footnote{ publicly available at https://github.com/hsu-aut/IndustrialStandard-SysML-Profile-VDI3682}.
The development of a profile for the VDI/VDE 3682 enables the guided modelling according to the standard and the seamless integration into the entire system development process. 
This increases the usability for the system engineer. 
The main advantage of using this profile instead of the classic \ac{uml} activity diagram is the reusability and the facilitated verification of the model during the modeling process. 
Reusability is achieved by using a standardized description of system functions. 
Model verification is achieved by defining \ac{ocl} constraints and is explained in Section \ref{sec:constraintmodeling}. 
\section{DSML for VDI/VDE 3682}
\label{sec:method}

\subsection{VDI/VDE 3682 Profile}
\label{sec:subprofile}
For the development of a \ac{dsml} for the VDI/VDE 3682, a lightweight approach based on \ac{sysml} is a suitable option.
The rationale behind this lies in the fact that elements from \ac{uml} and \ac{sysml} already offer analogous behavioral modeling constructs that readily lend themselves to straightforward extensions. 
Moreover, this approach facilitates the reuse of established tools, such as constraint verification through \ac{ocl}, and affords the seamless integration of these lightweight methodologies into extant \ac{mbse} workflows, such as \cite{beers2023mbse}. 
To achieve this, \ac{sysml} is extended with additional stereotypes to meet the requirements of VDI/VDE 3682. 
Therefore, this section introduces the elements of the \ac{dsml} and explains how they address the specific modeling requirements of this standard.
\autoref{tab:comparison} provides an overview of all elements in the \ac{dsml} along with their corresponding metaclasses, in comparison with the elements of the \ac{fpd}. 

\begin{table}[h]
\caption{Overview of the elements of the \ac{dsml}}

\label{tab:comparison}
\begin{tabular}{@{}p{0.25\linewidth} p{0.32\linewidth} p{0.34\linewidth}@{}}
\toprule
\textbf{\ac{fpd} Element} & \textbf{\ac{dsml} Element}& \textbf{\ac{sysml}/\ac{uml} Metaclass} \\ \midrule
FPD Diagram  & FPD Diagram & Activity Diagram\\ 
Process  & Process & Activity \\ 
System Boundary  & Diagram Boundary & - \\ 
Process Operator  & Process Operator & CallBehaviorAction \\ 
Technical Resource & Technical Resource & Class \\ 
Usage & Usage & Dependency \\ 
Product & Product & ActivityParameterNode\\ 
 & Intermediate Product &  ObjectNode\\ 
Energy & Energy & ActivityParameterNode\\
 & Intermediate Energy & ObjectNode\\ 
Information & Information & ActivityParameterNode\\ 
 & Intermediate Information & Object\-Node\\ 
Flow & Flow & ActivityFlow \\ 
 &  & ObjectFlow \\ 
Parallel Flow & ForkNode & ForkNode \\
 & JoinNode & JoinNode \\
Alternative Flow & MergeNode & MergeNode \\ 
 & DecisionNode & DecisionNode \\ 
Identification & Identification & Attribute \\
Characteristics & Characteristics & Attribute \\\bottomrule 
\end{tabular}
\end{table}
Following the VDI/VDE 3682 standard, every \fpb{Process} is defined by a single \fpb{System Boundary} (cf. \autoref{fig:fdp_classdia}), delineating the process from its external environment.   
This can be equated with the diagram boundary in the \ac{dsml}. 
The inputs and outputs of the system align with the elements situated on this diagram boundary, consistent with the standard. 

A process, according to the standard, can be understood as a description of a system's behavior.
Therefore it is appropriate to use a behavior diagram of \ac{sysml} as a fundamental diagram type. 
Particularly suitable for this purpose is an \sysml{Activity Diagram}, as within this diagram, both inputs and outputs can be described, and furthermore, decomposition of \fpb{Process Operators} are possible. 
Based on the \sysml{Activity Diagram}, a standalone diagram type was created, on which only the elements of the \ac{fpd} which are defined in the profile can be used. 
This leads to more guided modeling, helping to avoid modeling errors. 

Since each \fpb{\ac{fpd} Diagram} requires at least one \fpb{Process} element, a stereotype has been created for this element which extends the metaclass \sysml{Activity}. 
This element is always directly assigned to the \fpb{\ac{fpd} Diagram}. 
Additionally, a \sysml{Derived Property} has been created to ensure that the start and end \fpb{State} elements are displayed as additional \sysml{Attributes}.  
These attributes are subsequently used to verify whether each process has at least one start and one end \fpb{State}. 

\begin{figure}[htp]
    \centering
    \includegraphics[width=0.5\textwidth]{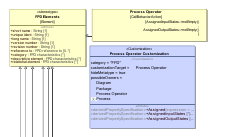}
   \caption{Stereotype and Customization of the Process Operator}
    \label{fig:snippetprofile}
\end{figure}

As shown in \autoref{fig:snippetprofile} the \fpb{Process Operator} is represented by a stereotype of the metaclass  \sysml{CallBehaviorAction}, which can be decomposed into another \ac{fpd} diagram. 
Within the element \sysml{Customization}, the \sysml{Possible Owners} are defined, restricting the usage of this stereotype. 
The derived properties refer to the \fpb{State}-describing elements entering or exiting the \fpb{Process Operator}, as well as to the \fpb{Technical Resource} linked to the \fpb{Process Operator} through the \fpb{Usage} relationship.
This \fpb{Usage} relationship is derived from the metaclass \sysml{Dependency}.
These are also utilized within the Constraints defined in \ac{ocl} to verify whether each \fpb{Process Operator} has at least one input and one output connected by a \fpb{flow}. 
This \fpb{Flow} is inherited from \sysml{ActivityFlow} and \sysml{ObjectFlow}, ensuring the representation of the sequential order and the proper passing of all information.
Through additional \sysml{Derived Properties}, it is ensured that all information can be passed on during the decomposition. 

The \fpb{Technical Resource} element is implemented through a corresponding stereotype, which belongs to the metaclass \sysml{Class} and also inherits the properties of the \ac{sysml} stereotype \sysml{System}. 
Since \ac{sysml} excludes the possibility of modeling elements outside diagram boundaries, the \fpb{Technical Resource} elements must be modeled within the \ac{fpd} diagram. 
This violates the standard, but it is not a serious breach.

The \fpb{State}-describing elements (\fpb{Product}, \fpb{Energy}, \fpb{Information}) that lie on the \fpb{System Boundary} have outputs only when they are introduced into the system or inputs only when they are leaving the system. 
However, the elements within the \fpb{System Boundary} must have both inputs and outputs.  
To formulate appropriate constraints for adhering to these rules, different stereotypes for the two variants have been modelled.
The \fpb{State}-describing elements on the \fpb{System Boundary}, therefore, receive the same name as those from the standard and are implemented as stereotypes of the metaclass \sysml{ActivityParameterNode},that are used at the beginning and end of activities to accept inputs and provide outputs. 
Those within the \fpb{System Boundary} receive the prefix Intermediate in their nomenclature and are implemented as stereotypes of the metaclass \sysml{ObjectNode}. 
This differentiation enables targeted querying and constraint modeling. 

The elements Identification and Characteristics are implemented using additional stereotypes, and all the subordinate \ac{fpd} elements are represented by \sysml{Attributes}. 
These elements are decomposed by nesting the corresponding attributes and stereotypes. 
In the profile, a parent stereotype called \fpb{\ac{fpd} Elements} has been introduced. 
All other \ac{fpd} stereotypes inherit the attributes of this stereotype through the generalization relationship. 

Within the \ac{fpd}, it is possible to model \fpb{parallel} and \fpb{alternative flows}. 
Parallel flows are implemented in the profile using the elements \sysml{ForkNode} and \sysml{JoinNode}, which are already part of the \ac{sysml} specification. 
Alternative flows are implemented through the \ac{sysml} elements \sysml{DecisionNode} and \sysml{MergeNode}. 
This approach indeed expands the modeling elements of the standard, but it allows for a meaningful grouping capability of input and output elements, as proposed in \cite{nabizada2022vorschlag}.

\subsection{Constraint Modeling}
\label{sec:constraintmodeling}
The meta model of the VDI/VDE 3682 imposes certain constraints on the use of its elements. 
Examples of these are: 
\begin{itemize}
    \item Each process must have at least one \fpb{Process Operator}. 
    \item The state-describing elements \fpb{Product}, \fpb{Information}, and \fpb{Energy} must always be associated with a \fpb{Process Operator}. 
    \item Each process must always have at least two state-describing elements.
\end{itemize}
These are reflected in the \ac{sysml} profile and can be verified through formal rules. 
However, certain aspects of the UML class diagram of the FPD standard are unclear.
For example, the cardinality of the relationship between \fpb{States} and processes indicates that a process must consist of at least two \fpb{State}-describing elements. 
Nevertheless, only the textual description clarifies that this involves at least one element describing the input and at least one element describing the output of the process. 

Additionally, it is useful for implementations to be able to distinguish whether the \fpb{State}-describing elements are located on or within the \fpb{system boundary}. 
Only in this way, formal rules can be defined to check whether the modeling rule regarding the connection of a \fpb{State}-describing element to a \fpb{Process Operator} is fulfilled. 
Through this approach, specific queries can be directed at the inputs and outputs of the corresponding elements to verify the existence of this connection. 
A \fpb{State}-describing element within the system boundary must have connections to a \fpb{Process Operator} as both an input and an output. 
For the \fpb{State}-describing elements on the system boundary, this connection must only be fulfilled either at the input or the output. 

These and additional constraints have been implemented using \ac{ocl} to enable standard-compliant modeling. 

\autoref{lst:ocl-flow} exemplifies a constraint written in \ac{ocl}. 
This constraint is associated with the \fpb{Flow} Element and defines that this element cannot simultaneously have an element of the \fpb{Information}, \fpb{Energy}, and \fpb{Product} stereotypes as both source and target.  
This constraint prevents two \fpb{State}-describing elements from being connected together. 
\begin{figure}[h]
\begin{lstlisting}[style=OCLStyle, caption=Constraint for Flow Stereotype defined in OCL, label=lst:ocl-flow]
context Flow inv FlowsSourceAndSource
((appliedStereotype->exists(name='Flow'))
and not (source->exists(out | out.oclIsTypeOf(Information) or out.oclIsTypeOf(Product) 
or out.oclIsTypeOf(IntermediateInformation)or out.oclIsTypeOf(IntermediateProduct)
or out.oclIsTypeOf(IntermediateEnergy)or out.oclIsTypeOf(Energy)) 
and target->exists(trg | trg.oclIsTypeOf(Information) or trg.oclIsTypeOf(Product)
or trg.oclIsTypeOf(IntermediateInformation)or trg.oclIsTypeOf(IntermediateProduct)
or trg.oclIsTypeOf(IntermediateEnergy) or trg.oclIsTypeOf(Energy))))
\end{lstlisting}
\end{figure}

\subsection{Automated Code Generation}
\label{sec:automatedcodegeneration}
However, \ac{xmi} is specifically tailored for exchanging metadata between modeling tools, such as \ac{sysml} modeling tools and other software engineering.
As highlighted in Section~\ref{sec:sota_3682approaches}, there is a need to utilize process descriptions in other description languages and interchange formats, such as \ac{aml} or \ac{owl}, in a standards-compliant manner. 
One approach to achieve this is the use of an established and versatile interchange format like ~\ac{xml}. 

The upcoming part 3 of the VDI/VDE 3682 envisions a unified \ac{xml} schema for representing the standard. 
A promising proposal for a standardized \ac{xml} representation has already been published with the approach outlined in \cite{nabizada2022vorschlag}. 
This approach is used here for the automatic generation of a formalized representation of the \ac{fpd} in \ac{xml}. 

To facilitate export into the \ac{xml} format, templates written in \ac{vtl} can be utilized. 
These templates enable the automatic generation of the interchange format from the existing \ac{sysml} model using the Apache Velocity Engine. 

\autoref{lst:vtl-snippet} displays a snippet from the \ac{vtl} code. 
The shown excerpt searches the Node attribute of all process elements.  
 If elements are found that exhibit the stereotypes \fpb{Product}, \fpb{Information}, or \fpb{Energy}, a \ac{xml} element \codein{state} with the corresponding attributes is generated for each of these elements.  
Additionally, references are made to the ID of the Process Operator linked with the \fpb{State}-describing Element and the IDs of the corresponding Flows. 
The template's defined structure facilitates precise mapping to the proposed \ac{xml} representation of the \ac{fpd} in \cite{nabizada2022vorschlag}. 
\begin{figure}[h]
\begin{lstlisting}[style=XMLStyle, caption=Snippet of the VTL Template, label=lst:vtl-snippet]
#foreach ($State in $Process.Node) 
#if($State.AppliedStereotype.size()>0)
#if($State.AppliedStereotype[0].Name == "Product" || $State.AppliedStereotype[0].Name == "Information" || $State.AppliedStereotype[0].Name == "Energy") 
            <state stateType="$State.AppliedStereotype[0].Name">
                <identification uniqueIdent="$State.ElementID" shortName="$State.ShortName" longName="$State.LongName" versionNumber="$State.versionNumber" revisionNumber="$State.revisionNumber">
                    <references></references>
                </identification>
                <characteristics>
                </characteristics>
                <assignments> 
#foreach($AssignedID in $State.assignedToProcessOperator) 
                    <assigned id="$AssignedID.ElementID" /> 
#end 
	        </assignments>
                <flows> 
#foreach($OutgoingFlow in $State.Outgoing) 
                    <flow>
                        <exit id="$OutgoingFlow.ElementID" />
                    </flow> #end
#foreach($IncomingFlow in $State.Incoming) 
                    <flow>
                        <entry id="$IncomingFlow.ElementID" />
\end{lstlisting}
\end{figure}

This export capability allows the approach to be utilized by users who want to model processes according to this standard independently of a Systems Engineering activity, such as for production capabilities \cite{kocher2020formal} or process parameter interdependencies \cite{tj2023}. 
Furthermore, this work takes another step towards standardizing an \ac{xml}-based exchange format for the VDI/VDE 3682, as it demonstrates the applicability of the proposal for mapping \ac{fpd} to \ac{xml} from \cite{nabizada2022vorschlag}. 

\section{Application Example}
\label{sec:applicationexample}
\begin{figure}[htp]
    \centering   \includegraphics[width=0.4\textwidth]{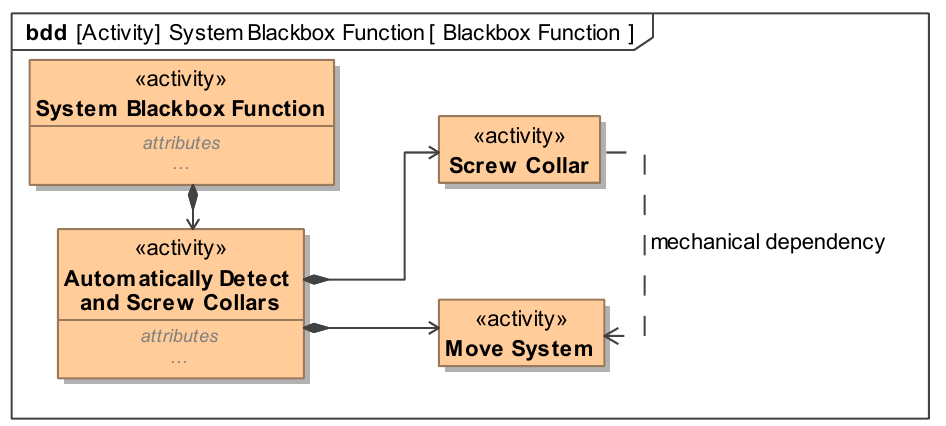}
    \caption{Blackbox functions of the Collar Screwing System}
    \label{fig:BBF}
\end{figure}

The profile mechanism introduced in Section~\ref{sec:method} is utilized and demonstrated using an application example from aircraft production. 
In this example, a system is developed to screw collars within an aircraft fuselage. 
The model was implemented using the \ac{msosa} software from Dassault Systèmes.
A workflow for the development of an aircraft production system has already been introduced in \cite{beers2023mbse}. 
This workflow is also used in this application example. 
During the development of the Functional Viewpoint, the Functional Requirements modeled in the Requirements Viewpoint are initially used to derive the Blackbox functions. 
An excerpt of the Blackbox functions can be found in \autoref{fig:BBF}. 
Building on this, the individual Blackbox functions are intended to be connected, and a sequential series of steps is defined. 
For this purpose, the newly developed profile is  applied.  

\begin{figure}[htp]
    \centering   \includegraphics[width=0.5\textwidth]{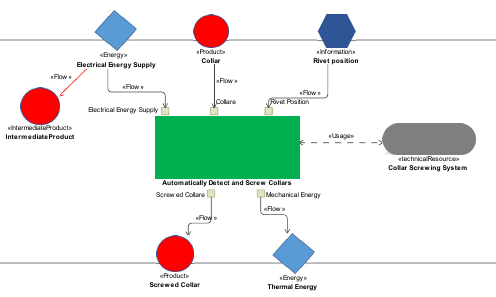}
    \caption{Implementation Example}
    \label{fig:fpd_applicationexample}
\end{figure}

The \sysml{Activity} elements of the Blackbox functions are assigned to the stereotype \fpb{Process}. 
Subsequently, an \fpb{\ac{fpd} Diagram} is created for each of these processes.  
This is exemplified in \autoref{fig:fpd_applicationexample}. 

The initial \fpb{State} consists of the \fpb{Product} \textit{Collar} and the \fpb{Information} \textit{Rivet Position}. 
Additionally, an energy supply must be ensured during the process. 
The \fpb{Process Operator} \textit{Automated Collar Screwing} transforms this initial State into the end State \textit{Screwed Collar} as the \fpb{Product} and \textit{Thermal Energy} as the \fpb{Energy} output. 
The inputs and outputs of the system functions can be derived from the system context. 
A detailed insight into this aspect can be found in \cite{beers2023mbse}. 

Subsequently, the \fpb{Process Operator} is decomposed into additional sub-processes. 
In \autoref{fig:fpd_applicationexample}, an error was deliberately introduced to exemplify error identification.  
In the example, two state-describing elements were connected (see red arrow). 
While the \fpb{Flow} stereotype is permitted to have a \fpb{State} stereotype as both source and target types, the \ac{ocl} constraints prohibit this modeling. 
The user is alerted to the corresponding error through the error message: \enquote{A state must always be assigned a process operator. Linking two states is not permitted}.

After addressing all errors, the process description can be automatically exported to \ac{xml} following the approach described in Section \ref{sec:automatedcodegeneration} and using the internal report wizard of \ac{msosa}. 
An excerpt of the serialization is exemplified in \autoref{lst:xml_serial} and demonstrates the mapping of the input elements from \autoref{fig:fpd_applicationexample} to the XML schema proposed by \cite{nabizada2022vorschlag}. 
For clarity, the listing does not display any Characteristics, but they can be created through the profile as mentioned in Section \ref{sec:subprofile}. 
Nevertheless, it can be seen that the process model has been transferred to the correct exchange format. 
\begin{figure}
\begin{lstlisting}[style=XMLStyle, caption=Excerpt of the XML serialisation, label=lst:xml_serial]
<process id="_2021x_56901f2_1698910860901_408207_17462"> 
    <systemLimit id="_2021x_56901f2_1698910860895_144044_17461" shortName="Modeling Example" />
        <states> 
            <state stateType="Energy">
                <identification uniqueIdent="_2021x_56901f2_1698911485391_499621_17572" shortName="Electrical Energy Supply" longName="" versionNumber="" revisionNumber="">
                    <references></references>
                </identification>
                <characteristics>
                </characteristics>
                <assignments> 
                    <assigned id="_2021x_56901f2_1698911444248_32009_17539" /> 
                </assignments>
                <flows> 
                    <flow>
                        <exit id="_2021x_56901f2_1698911552130_199686_17659" />
                    </flow>                     
                </flows>
            </state>
            <state stateType="Product">
                <identification uniqueIdent="_2021x_56901f2_1698911586000_254919_17680" shortName="Collar" longName="" versionNumber="" revisionNumber="">
                    <references></references>
                </identification>
                <characteristics>
                </characteristics>
                <assignments> 
                    <assigned id="_2021x_56901f2_1698911444248_32009_17539" /> 
                </assignments>
                <flows> 
                    <flow>
                        <exit id="_2021x_56901f2_1698911938863_856843_17815" />
                    </flow>                 
                </flows>
            </state>
            <state stateType="Information">
                <identification uniqueIdent="_2021x_56901f2_1698911694625_303176_17766" shortName="Rivet position" longName="" versionNumber="" revisionNumber="">
                    ...
\end{lstlisting}
\end{figure}

\section{Summary and Outlook}
\label{sec:summaryoutlook}
By using the \ac{fpd} according to VDI/VDE 3682, processes can be structured and modeled using simple means based on the \ac{ppr} concept. 
This proves to be particularly effective in Systems Engineering as an instrument to break down the functional Blackbox and transition to a functional Whitebox. 

In this context, the utilization of a \ac{dsml} was motivated and developed as a profile of \ac{sysml}.
In addition to graphical modeling capabilities, this \ac{dsml} incorporates \ac{ocl} rules, allowing for the verification of modeling errors and inconsistencies. 
The \ac{dsml} enables process modeling according to the standard within \ac{mbse} tools. 
The process descriptions can be not only utilized within \ac{mbse} tools but also serialized in XML, enabling their application in various contexts. 
Finally, using a case study from aircraft production  \cite{gehlhoff2022challenges}, it was demonstrated how the \ac{dsml} facilitates the transition from a Blackbox to a Whitebox. 

Future work will be concerned with automating several mechanisms and integrating them into a plug-in using the \ac{dsml}. 
This will enhance the guidance and support for process modeling. 
Additionally, the approach is intended to be applied to further use cases in Systems Engineering to further improve it and show its general applicability.
The process modeling is used within the Functional Viewpoint, which was presented in the paper \cite{beers2023mbse} and is based on the SPES method. 
However, the representation of VDI/VDE 3682 does not only include the process. 
It also includes the \fpb{Technical Resource}.
The \fpb{Technical Resources} described there can be used as elements of the Logical Viewpoint to describe the logical architecture of the system to be developed. 
These will be further specified in the Technical Viewpoint later on. 
How to appropriately model the technical resources will be presented in a future paper. 
\\
\\
\\
\bibliographystyle{IEEEtran}
\bibliography{references}

\begin{thebibliography}{10}
\providecommand{\url}[1]{#1}
\csname url@samestyle\endcsname
\providecommand{\newblock}{\relax}
\providecommand{\bibinfo}[2]{#2}
\providecommand{\BIBentrySTDinterwordspacing}{\spaceskip=0pt\relax}
\providecommand{\BIBentryALTinterwordstretchfactor}{4}
\providecommand{\BIBentryALTinterwordspacing}{\spaceskip=\fontdimen2\font plus
\BIBentryALTinterwordstretchfactor\fontdimen3\font minus \fontdimen4\font\relax}
\providecommand{\BIBforeignlanguage}[2]{{%
\expandafter\ifx\csname l@#1\endcsname\relax
\typeout{** WARNING: IEEEtran.bst: No hyphenation pattern has been}%
\typeout{** loaded for the language `#1'. Using the pattern for}%
\typeout{** the default language instead.}%
\else
\language=\csname l@#1\endcsname
\fi
#2}}
\providecommand{\BIBdecl}{\relax}
\BIBdecl

\bibitem{d2017systems}
J.~D'Ambrosio and G.~Soremekun, ``{Systems engineering challenges and MBSE opportunities for automotive system design},'' in \emph{2017 IEEE International Conference on Systems, Man, and Cybernetics (SMC)}, 2017, pp. 2075--2080.

\bibitem{sanfilippo2018modeling}
E.~M. Sanfilippo, S.~Benavent, S.~Borgo, N.~Guarino, N.~Troquard, F.~Romero, P.~Rosado, L.~Solano, F.~Belkadi, and A.~Bernard, ``{Modeling Manufacturing Resources: An Ontological Approach},'' in \emph{Product Lifecycle Management to Support Industry 4.0}, 2018, pp. 304--313.

\bibitem{pohl2012model}
K.~Pohl, H.~H{\"o}nninger, R.~Achatz, and M.~Broy, \emph{{Model-based engineering of embedded systems: The SPES 2020 methodology}}.\hskip 1em plus 0.5em minus 0.4em\relax Springer, 2012.

\bibitem{VDIVDEGesellschaftMessundAutomatisierungstechnik.05.2015}
{VDI/VDE 3682-1}, ``{Formalised Process Descriptions - Concept and Graphic Representation},'' 2015.

\bibitem{gehlhoff2022challenges}
F.~Gehlhoff, H.~Nabizada, M.~Weigand, L.~Beers, O.~Ismail, A.~Wenzel, A.~Fay, P.~Nyhuis, W.~Lagutin, and M.~R{\"o}hrig, ``Challenges in automated commercial aircraft production,'' \emph{IFAC-PapersOnLine}, vol.~55, no.~2, pp. 354--359, 2022.

\bibitem{VDIVDEGesellschaftMessundAutomatisierungstechnik.05.2015.part2}
{VDI/VDE 3682-2}, ``{Formalised Process Descriptions - Information model},'' 2015.

\bibitem{frank2013domain}
U.~Frank, ``{Domain-Specific Modeling Languages: Requirements Analysis and Design Guidelines},'' \emph{{Domain Engineering: Product Lines, Languages, and Conceptual Models}}, pp. 133--157, 2013.

\bibitem{lagarde2008leveraging}
F.~Lagarde, H.~Espinoza, F.~Terrier, C.~Andr{\'e}, and S.~G{\'e}rard, ``{Leveraging Patterns on Domain Models to Improve UML Profile Definition},'' in \emph{International Conference on Fundamental Approaches to Software Engineering}.\hskip 1em plus 0.5em minus 0.4em\relax Springer, 2008, pp. 116--130.

\bibitem{seidl2012uml}
M.~Seidl, M.~Scholz, C.~Huemer, and G.~Kappel, \emph{{UML @ Classroom: An Introduction to Object-Oriented Modeling}}.\hskip 1em plus 0.5em minus 0.4em\relax Springer, 2015.

\bibitem{nomagic.2015}
\BIBentryALTinterwordspacing
{MagicDraw}, ``{UML Profiling and DSL 18.1 - User Guide},'' 2015. [Online]. Available: \url{https://docs.nomagic.com/download/attachments/17667876/MagicDraw%20UMLProfiling%26DSL%20UserGuide.pdf}
\BIBentrySTDinterwordspacing

\bibitem{bousse2012aligning}
E.~Bousse, D.~Mentr{\'e}, B.~Combemale, B.~Baudry, and T.~Katsuragi, ``Aligning sysml with the b method to provide v\&v for systems engineering,'' in \emph{Proceedings of the workshop on model-driven engineering, verification and validation}, 2012, pp. 11--16.

\bibitem{cabot2012object}
J.~Cabot and M.~Gogolla, ``{Object Constraint Language (OCL): A Definitive Guide},'' in \emph{Formal Methods for Model-Driven Engineering: 12th International School on Formal Methods for the Design of Computer, Communication, and Software Systems, SFM 2012}.\hskip 1em plus 0.5em minus 0.4em\relax Springer, 2012, pp. 58--90.

\bibitem{hildebrandt2018ontology}
C.~Hildebrandt, S.~T{\"o}rsleff, B.~Caesar, and A.~Fay, ``Ontology building for cyber-physical systems: A domain expert-centric approach,'' in \emph{2018 IEEE 14th international conference on automation science and engineering (CASE)}.\hskip 1em plus 0.5em minus 0.4em\relax IEEE, 2018, pp. 1079--1086.

\bibitem{drath2021automationml}
R.~Drath, \emph{{AutomationML: A Practical Guide}}.\hskip 1em plus 0.5em minus 0.4em\relax Walter de Gruyter GmbH \& Co KG, 2021.

\bibitem{jager2012durchgangige}
T.~J{\"a}ger, L.~Christiansen, M.~Strube, and A.~Fay, ``{Durchg{\"a}ngige Werkzeugunterst{\"u}tzung von der Anforderungserhebung bis zur Anlagenstrukturbeschreibung mittels formalisierter Prozessbeschreibung und AutomationML},'' \emph{Proceedings of EKA}, pp. 8--10, 2012.

\bibitem{kocher2020formal}
A.~K{\"o}cher, C.~Hildebrandt, L.~M. Vieira~da Silva, and A.~Fay, ``{A Formal Capability and Skill Model for Use in Plug and Produce Scenarios},'' in \emph{2020 25th IEEE International Conference on Emerging Technologies and Factory Automation (ETFA)}, vol.~1, 2020, pp. 1663--1670.

\bibitem{csspaper}
A.~Köcher, A.~Belyaev, J.~Hermann, J.~Bock, K.~Meixner, M.~Volkmann, M.~Winter, P.~Zimmermann, S.~Grimm, and C.~Diedrich, ``A reference model for common understanding of capabilities and skills in manufacturing,'' \emph{at - Automatisierungstechnik}, vol.~71, no.~2, pp. 94--104, 2023.

\bibitem{caesar2020information}
B.~Caesar, A.~H{\"a}nel, E.~Wenkler, C.~Corinth, S.~Ihlenfeldt, and A.~Fay, ``{Information Model of a Digital Process Twin for Machining Processes},'' in \emph{{2020 25th IEEE International Conference on Emerging Technologies and Factory Automation (ETFA)}}, vol.~1, 2020, pp. 1765--1772.

\bibitem{tj2023}
T.~Jeleniewski, H.~Nabizada, J.~Reif, A.~Köcher, and A.~Fay, ``{A Semantic Model to Express Process Parameters and their Interdependencies in Manufacturing},'' in \emph{2023 IEEE 32nd International Symposium on Industrial Electronics (ISIE)}, 2023, pp. 1--6.

\bibitem{kathrein2019meta}
L.~Kathrein, K.~Meixner, D.~Winkler, A.~L{\"u}der, and S.~Biffl, ``{A Meta-Model for Representing Consistency as Extension to the Formal Process Description},'' in \emph{2019 24th IEEE International Conference on Emerging Technologies and Factory Automation (ETFA)}, 2019, pp. 1653--1656.

\bibitem{novak2022digitalized}
P.~Nov{\'a}k and J.~Vysko{\v{c}}il, ``{Digitalized Automation Engineering of Industry 4.0 Production Systems and Their Tight Cooperation with Digital Twins},'' \emph{Processes}, vol.~10, no.~2, p. 404, 2022.

\bibitem{nabizada2020offenes}
H.~Nabizada, A.~K{\"o}cher, C.~Hildebrandt, and A.~Fay, ``{Offenes, webbasiertes Werkzeug zur Informationsmodellierung mit Formalisierter Prozessbeschreibung},'' in \emph{Automation 2020 - Shaping Automation for our Future}, 2020.

\bibitem{nabizada2022vorschlag}
H.~Nabizada, T.~Jeleniewski, A.~K{\"o}cher, and A.~Fay, ``{Vorschlag f{\"u}r eine XML-Repr{\"a}sentation der Formalisierten Prozessbeschreibung nach VDI/VDE 3682},'' in \emph{{17. Fachtagung EKA - Entwurf komplexer Automatisierungssysteme}}, 2022.

\bibitem{ObjectManagementGroup.2019}
\BIBentryALTinterwordspacing
{OMG}, ``{Systems Modeling Language (SysML{\texttrademark} 1.6)},'' 2019. [Online]. Available: \url{https://www.omg.org/spec/SysML/1.6/PDF}
\BIBentrySTDinterwordspacing

\bibitem{pohl2016advanced}
K.~Pohl, M.~Broy, H.~Daembkes, and H.~H{\"o}nninger, \emph{{Advanced Model-Based Engineering of Embedded Systems: Extensions of the SPES 2020 Methodology}}.\hskip 1em plus 0.5em minus 0.4em\relax Springer, 2016.

\bibitem{bohm2021model}
W.~B{\"o}hm, M.~Broy, C.~Klein, K.~Pohl, B.~Rumpe, and S.~Schr{\"o}ck, \emph{{Model-Based Engineering of Collaborative Embedded Systems: Extensions of the SPES Methodology}}.\hskip 1em plus 0.5em minus 0.4em\relax Springer Nature, 2021.

\bibitem{hayward2022sysml}
A.~Hayward, M.~Rappl, and A.~Fay, ``{A SysML-based Function-Centered Approach for the Modeling of System Groups for Collaborative Cyber-Physical Systems},'' in \emph{2022 IEEE International Systems Conference (SysCon)}, 2022, pp. 1--8.

\bibitem{beers2023mbse}
L.~Beers, M.~Weigand, H.~Nabizada, and A.~Fay, ``{MBSE Modeling Workflow for the Development of Automated Aircraft Production Systems},'' in \emph{2023 IEEE 28th International Conference on Emerging Technologies and Factory Automation (ETFA)}, 2023, pp. 1--8.

\bibitem{weilkiens2022model}
T.~Weilkiens, J.~G. Lamm, S.~Roth, and M.~Walker, \emph{{Model-Based System Architecture}}.\hskip 1em plus 0.5em minus 0.4em\relax John Wiley \& Sons, 2022.

\bibitem{binder2023towards}
C.~Binder, P.~H{\"u}necke, C.~Neureiter, and A.~L{\"u}der, ``{Towards flexible production systems engineering according to RAMI 4.0 by utilizing PPR notation},'' in \emph{2023 IEEE 21st International Conference on Industrial Informatics (INDIN)}, 2023, pp. 1--6.

\end{thebibliography}

\end{document}